\documentclass[a4paper,12pt]{article}
\usepackage[utf8]{inputenc}
\usepackage{amssymb}
\usepackage{hhline}
\usepackage{ifthen}
\usepackage{hyperref}

\newenvironment{proof}[1][\hspace{-1.0ex}]%
{\par\addvspace{1mm}{\sc Proof\hspace{1.0ex}{#1}.} }%
{\quad$\blacktriangle$\par\addvspace{1mm}}

\newif\ifNoRemark
\def\addtheorem#1#2#3#4{
\ifthenelse{\equal{#2}{}}{}%
{\ifthenelse{\expandafter\isundefined\csname the#2\endcsname}{\newcounter{#2}}{}}
\newenvironment{#1}[1][\global\NoRemarktrue]
{\par\addvspace{2mm plus 0.5mm minus 0.2mm}\noindent 
{\bf #3}\ifthenelse{\equal{#2}{}}{}%
{\refstepcounter{#2}{\bf ~\csname the#2\endcsname}}%
{\bf \vphantom{##1}\ifNoRemark.\ \else\ (##1).\fi}\begingroup #4}%
{\endgroup\par\addvspace{1mm plus 0.5mm minus 0.2mm}\global\NoRemarkfalse}
\expandafter\newcommand\csname b#1\endcsname{\begin{#1}}
\expandafter\newcommand\csname e#1\endcsname{\end{#1}}
}

\addtheorem{remark}{}{Remark}{\rm}
\addtheorem{theorem}{}{Theorem}{\sl}
\addtheorem{lemma}{lmm}{Lemma}{\sl}
\addtheorem{proposition}{prpstn}{Proposition}{\sl}

\title{A partition of the hypercube 
\\ into 
maximally 
nonparallel Hamming codes%
\thanks{This is the peer reviewed version of the following article: 
[D. S. Krotov. A partition of the hypercube 
into maximally nonparallel Hamming codes. 
J. Comb. Des. 24(4) 2014, 179-187], 
which has been published in final form at 
\url{http://dx.doi.org/10.1002/jcd.21363}. 
This article may be used for non-commercial purposes in accordance with 
Wiley Terms and Conditions for Self-Archiving.}%
}
\author{Denis S. Krotov\thanks{Sobolev Institute of Mathematics, Novosibirsk, Russia; 
Novosibirsk State University, Novosibirsk, Russia.
E-mail: \texttt{krotov@math.nsc.ru}}}

\begin{document}

\maketitle

\begin{abstract}
By using the Gold map, 
we construct a partition of the hypercube 
into cosets of Hamming codes 
such that for every two cosets 
the corresponding Hamming codes are maximally nonparallel, that is,
their intersection cardinality is as small as possible 
to admit nonintersecting cosets.
\end{abstract}

\section{Introduction}\label{s:intro}
Partitions of the hypercube into $1$-perfect codes, 
especially into $1$-perfect codes 
that are not translations of each other, 
attract some attention in literature 
\cite{HedSol:2009,Phelps:2000,Sol:2009:part,SolGus:2011:part}.
Such partitions themselves are perfect codes 
in a mixed alphabet, 
and they are used in the construction 
of binary perfect codes.
In this note, we apply crooked permutations 
in order to construct a class of such partitions 
with extremal properties. 
The partitions constructed consist 
of cosets of linear codes. 
On the other hand, such a partition is, 
in some sense, as far as possible from being `linear':
the affine span of every two of its codes 
coincides with the whole space.
The result of the note 
provides one more example of application 
of the crooked functions in the construction
of extremal combinatorial structures. 
In the main part of the note
(Section~\ref{s:constr} and~\ref{s:auto})
 we use the Gold functions for construction.
 Section~\ref{s:constr} describes the construction, and 
in Section~\ref{s:auto} the symmetries 
of the constructed partitions are considered.
The crooked permutations in general 
are briefly discussed in Section~\ref{s:qr}. 

We conclude the introduction with a brief survey 
of the results on $1$-perfect partitions, which was suggested to be included 
by one of the reviewers.

\subsection*{Survey: $1$-perfect partitions}
There are several works on constructions of $1$-perfect partitions 
(partitions of the hypercube into $1$-perfect codes, 
i.e., codes with parameters of the Hamming code)
and some papers where such partitions are used 
in the construction of $1$-perfect codes. 

The first construction of $1$-perfect codes that uses $1$-perfect partitions 
was proposed by Solov'eva in \cite{Sol:81} 
and, independently, by Phelps in \cite{Phelps:83}.
The construction uses two $1$-perfect partitions of length $n$ 
to construct a $1$-perfect code of length $2n+1$.
In the subsequent paper \cite{Phelps84}, Phelps generalized the construction: 
$2^k$ $1$-perfect partitions are combined 
to construct a $1$-perfect code of length $2^k(n+1)-1$.
It should be noted that the earlier construction by Heden \cite{Heden:77}, 
which was formulated in terms of codes over mixed alphabet 
instead of partitions,
is equivalent to the Solov'eva construction with some restriction: 
one of the partitions consists of translations of the same code.
The constructions of Solov'eva \cite{Sol:81} 
and Phelps \cite{Phelps84} exploit the principles
of the $X4$ construction \cite[18.7.2]{MWS} 
and the generalized concatenation by Zinoviev
\cite{Zin1976:GCC}, \cite[18.8.2]{MWS}, respectively, 
but also add some more variability
to the construction, which is important 
for constructing $1$-perfect codes
with different properties
and was used in further research by different authors. 
A different application of the generalized concatenation 
was proposed by Zinoviev in \cite{Zinoviev:88thesis} for $1$-perfect codes: 
$N$ $1$-perfect partitions of length $n$ 
and a $1$-perfect code of length $N$ over the $(n+1)$-ary alphabet 
(while $1$-perfect codes over non-binary alphabets are not considered in this paper, 
this is the only place where they are mentioned)
can be combined to construct a $1$-perfect code of length $nN$.

In \cite{Sol:81}, Solov'eva proposed 
two constructions of $1$-perfect partitions,
one of them 
(an analog of the known Vasil'ev construction 
of $1$-perfect codes)
giving at least $2^{2^{(n-1)/2}}$ 
different partitions.

Rif\`a and Vardy \cite{RV:97:partitions} 
proved that for every $1$-perfect code, 
there are nonequivalent $1$-perfect partitions 
that contain this code as a cell.

Rif\`a \cite{Rifa:99:STS} 
constructed $1$-perfect partitions
in an algebraic way, 
extending Steiner triple systems (STS) 
that satisfy some special properties, 
so-called well-ordered STS.
Rif\`a, Pujol and Borges in \cite{RPB:2001:partitions}, 
Borges, Fernandez, Rif\`a and Villanueva in \cite{BFRV:2001:partitions} 
established connections between
some combinatorial and metrical properties of a $1$-perfect partition 
and algebraic properties of its codes.

In \cite{Phelps:2000}, Phelps enumerated all $11$ nonequivalent $1$-perfect
partitions of the $7$-cube and used them to evaluate the number 
of nonequivalent $1$-perfect codes of length $15$ 
(the exact number was found nine years later by \"Osterg\aa rd and Pottonen 
\cite{OstPot:15}).

In \cite{ALS:2001}, Avgustinovich, Lobstein, and Solov'eva investigated
the arrays of the cardinalities of the pairwise intersections 
of the cells of two $1$-perfect partitions.

In \cite{ASH:2003:part},
by Avgustinovich, Solov'eva, and Heden,
partitions of the space into mutually nonequivalent 
$1$-perfect codes are constructed for every length $2^m-1\geq 32$.

In \cite{HedSol:2009}, 
Heden and Solov'eva constructed 
several classes of partitions of the space 
into mutually nonparallel cosets of Hamming codes.
Solov'eva \cite{Sol:2009:part}, 
Solov'eva and Gus'kov \cite{SolGus:2011:part}
considered the construction of cell-transitive, 
$2$-cell-transitive, and vertex transitive $1$-perfect partitions 
(see the definitions in Section~\ref{s:auto}),
including partitions into mutually nonparallel cosets 
of Hamming codes \cite{Sol:2009:part},
by using known recursive constructions of $1$-perfect codes.
All known general constructions \cite{HedSol:2009}, \cite{Sol:2009:part}
of partitions of the space into cosets of Hamming codes 
give partitions in which some of the cosets 
are not maximally nonparallel 
in the sense considered in this paper. 
The only exception is the partition (Partition 8) 
of length $7$ found in \cite{Phelps:2000}.
Moreover, 
for all these constructions,
the `non-parallelity index' of a partition $ \{C_0,\ldots,C_n\}$
defined as
$$
\min_{i,j,i\neq j} ( \mathrm{dim}\langle C_i\cup C_j \rangle - \mathrm{dim}\, C_i )
$$
do not increase with the growth of the length $n$.
For the partition constructed in Section~\ref{s:constr}, 
this index has the order $\log n$,
which implies an essential improvement in the direction of `non-parallelity'.

\section{Construction of the partition} \label{s:constr}
Let $m\geq 3$ be odd and let $F$ be the finite field $\mathrm{GF}(2^m)$ of order $2^m$.
Let $\sigma$ be a power of $2$, and assume that $\sigma \pm 1$ and $2^m-1$ are relatively prime,
which is, by simple arguments, equivalent to the condition $\mathrm{gcd}(s,m)=1$, where $\sigma = 2^s$. 
For example, $\sigma=2$.
Since the powers of a primitive element generating $F^*=F\setminus \{0\}$ 
are calculated modulo $2^m-1$,
the condition $\mathrm{gcd}(\sigma \pm 1,2^m-1)=1$ means that 
both $x\to x^{\sigma+1}$, which is known as the Gold map, 
and $x\to x^{\sigma-1}$ are one-to-one mappings.
We will treat the codes $C$ of length $2^m$ (or $2^m-1$) 
as collections of subsets of $F$ 
($F^*=F\setminus \{0\}$, respectively), 
i.e., $C\subset 2^F$ ($C\subset 2^{F^*}$). 
The \emph{code distance} is the minimal cardinality of the symmetric difference 
$X\triangle Y = (X \setminus Y) \cup (Y \setminus X)$ 
of two different elements $X$, $Y$ of the code. 
A code is \emph{linear} if it is a subspace of the vector space 
$(2^F,\triangle, \cdot)$ over $\mathrm{GF}(2)$
(the set $2^F$ of all subsets of $F$ is endowed 
with the addition $\triangle$ and the natural multiplication by a scalar:
$1\cdot X= X$, $0\cdot X = \emptyset$).
A \emph{Hamming code} (\emph{extended Hamming code}) 
is defined as a collection of subsets $X$ of $F^*$
(even-cardinality subsets of $F$, respectively)
satisfying $ \sum_{x\in X} \pi(x) = 0 $ 
where $\pi$ is some fixed permutation of $F^*$ 
($F$, respectively).

Recall some facts:

\begin{enumerate}
\item[(A)] for all $x,y\in F$: $(x+y)^\sigma = x^\sigma+y^\sigma$ 
(derived from $(x+y)^2=x^2+y^2$);

\item[(B)] for all $x\in F$: $x^{\sigma}+x+1 \neq 0$ (indeed, otherwise 
$(x+1)^{\sigma+1}=(x+1)(x+1)^{\sigma}=(x+1)(x^{\sigma}+1)=x^{\sigma+1}+x^{\sigma}+x+1=x^{\sigma+1}$,
which is impossible as $f(x)=x^{\sigma+1}$ is one-to-one);

\item[(C)] the cardinality of the code $B=\{ X\in 2^F \, : \: \sum_{x\in X} 1 = 0, \ \sum_{x\in X} x = 0, \ \sum_{x\in X} x^{\sigma+1} = 0\}$ is $2^{2^m-2m-1}$
(in fact, $B$ is a linear code 
of distance at least $6$,
which has the maximal cardinality among 
the linear distance-$6$ codes of length $2^m$ 
\cite{CarChaZin}; 
in the case $\sigma=2$, a BCH code).
\end{enumerate}

For $\alpha \in F$, $p\in \{0,1\}$,
define the code $H_\alpha^p$ as the collection of subsets $X$ of $F$ satisfying
\begin{eqnarray*}
&&\sum_{x\in X} 1 = p, \\
&&\sum_{x\in X} (x+\alpha)^{\sigma+1} = 0. 
\end{eqnarray*}
Clearly, $H_\alpha^0$ is an extended Hamming code, and $H_\alpha^1$ is a coset of $H_\alpha^0$.

\begin{theorem} 
{\rm (i)} 
  The codes $H_\alpha^1$, $\alpha\in F$, 
  are mutually disjoint.
{\rm (ii)} 
  The intersection of two different codes 
  $H_\alpha^0$ and $H_\beta^0$ 
  has the  cardinality $2^{2^m-2m}$.
\end{theorem}

\begin{proof}
An element $X$ of the intersection of $H_\alpha^p$ and $H_\beta^p$ satisfies 
\begin{eqnarray}
&&\sum_{x+\alpha\in X} 1  = p , \label{eq:1}\\
&&\sum_{x+\alpha\in X} x^{\sigma+1}  = 0 , \label{eq:2}\\
&&\sum_{x+\alpha\in X} (x+\alpha+\beta)^{\sigma+1}  = 0 . \label{eq:3}
\end{eqnarray}
We derive
\begin{eqnarray*}
 && \sum_{x+\alpha\in X} (x+\alpha+\beta)^{\sigma+1}  
= \sum_{x+\alpha\in X} (x+(\alpha+\beta))^{\sigma}(x+(\alpha+\beta)) \\
&\stackrel{\rm(A)}{=}& \sum_{x+\alpha\in X} x^{\sigma+1} + \sum_{x+\alpha\in X} x^{\sigma} (\alpha+\beta) + \sum_{x+\alpha\in X} x (\alpha+\beta)^{\sigma} + \sum_{x+\alpha\in X}(\alpha+\beta)^{\sigma+1}\\
&\stackrel{(\ref{eq:2}){\rm(A)}(\ref{eq:1})}{=}& \Big(\sum_{x+\alpha\in X} x\Big)^{\sigma} (\alpha+\beta) + \Big(\sum_{x+\alpha\in X} x\Big) (\alpha+\beta)^{\sigma} + p(\alpha+\beta)^{\sigma+1}.
\end{eqnarray*}
For  $p=1$ and $\alpha \neq \beta$, the last expression cannot be equal to $0$, by (B), 
which contradicts (\ref{eq:3}) and proves (i).

For $p=0$, the last expression above is equal to 
$$ \Big(\sum_{x+\alpha\in X} x\Big) \Bigg( \Big(\sum_{x+\alpha\in X} x\Big)^{\sigma-1} + (\alpha+\beta)^{\sigma-1} \Bigg)(\alpha+\beta),$$
which implies, together with (\ref{eq:3}) and $\alpha \neq \beta$, that 
$$ \mbox{either}\ \  \sum_{x+\alpha\in X} x = 0 \qquad\mbox{or}\qquad \sum_{x+\alpha\in X} x = \alpha+\beta. $$
By (C), each of these two cases,  
together with (\ref{eq:1}) and (\ref{eq:2}), 
has exactly $2^{2^m-2m-1}$ solutions for $X$. 
So, there are $2^{2^m-2m}$ solutions in total, which proves (ii).
\end{proof}

As a \textbf{corollary}, we partitioned all the odd-cardinality subsets of $F$ 
into $2^m$ cosets $H_\alpha^1$ of extended Hamming codes such that every two different cosets are maximally nonparallel. Being maximally nonparallel here means that 
the intersection of the corresponding extended Hamming codes is as small as possible 
to admit disjoint odd cosets. 
Note that, in general, the dimension of the intersection of two Hamming codes (extended Hamming codes) 
can  possess any value from ${2^m-m-1}$ to ${2^m-2m-1}$ \cite[Theorem 3.4]{EV:98}. 
But in the last case, the linear span of the two codes will coincide with the whole space 
(in the case of extended Hamming codes, with the set of even-cardinality subsets of $F$),
which means that the cosets of these codes (odd cosets, for extended Hamming codes) necessarily intersect.

By removing the zero element from all $X$, we obtain a partition of the $(2^m-1)$-cube into 
maximally nonparallel cosets of Hamming codes.
The first ($m=3$) partition is equivalent to Partition 8 in the classification 
of all partitions of the $7$-cube into $1$-perfect codes given by Phelps in \cite[Appendix]{Phelps:2000}.

\textbf{Remark.} We can consider $\alpha$ as the ``color'' of the elements of $H_\alpha^1$. 
It is easy to see that, given an odd-cardinality set $X\subset F$, 
its color can be explicitly calculated by the formula
$$
\alpha= \alpha(X) = \sum_{x\in X}x + \Big(\sum_{x,y\in X,\, x\neq y} x^\sigma y\Big)^{1/(\sigma+1)}. 
$$

\section{Automorphisms and orbits}  \label{s:auto}
Let us consider some isometries of the space 
that stabilize the constructed partition 
$\mathbf{H}=\{H_\alpha^1\}_{\alpha\in F}$
of the odd-cardinality subsets of $F$. 
For convenience, define 
$H_\alpha(\beta)$, $\alpha,\beta\in F$, 
as the set of odd-cardinality 
subsets $X\subset F$ satisfying
\begin{equation}\label{eq:Hab}
 \sum_{x\in X} (x+\alpha)^{\sigma+1} = \beta^{\sigma+1}
\end{equation}
(in particular, $H_\alpha(0)=H_\alpha^1$), 
and define $H$ as the set 
of even-cardinality subsets $Y\subset F$ satisfying
\begin{equation}\label{eq:H}
\sum_{x\in Y} x = 0
\end{equation}
(recall that $H$ is an extended Hamming code; 
so, every odd-cardinality subset of $F$ 
is at distance one from exactly one element of $H$).

Direct verifications confirm the validity 
of the following four statements:

\begin{lemma}\label{l1} 
For every $\delta$ from $F$, 
the permutation $x\to x+\delta$ of $F$ 
maps $H_\alpha(\beta)$ 
to $H_{\alpha+\delta}(\beta)$.
\end{lemma}
\begin{proof}
If $X$ meets (\ref{eq:Hab}), 
then $Y=X+\delta$ satisfies
$\sum_{x\in Y+\delta} (x+\alpha)^{\sigma+1} = \beta^{\sigma+1},$
which is equivalent to 
$\sum_{y\in Y} (y+(\alpha + \delta))^{\sigma+1} = \beta^{\sigma+1}.$
\end{proof}  

\begin{lemma}\label{l2} 
For every $\mu$ from $F^*$, 
the permutation $x\to \mu x$ of $F$ 
maps $H_\alpha(\beta)$ 
to $H_{\mu\alpha}(\mu\beta)$.
\end{lemma}
\begin{proof}
If $X$ meets (\ref{eq:Hab}), 
then $Y=\mu X$ satisfies
$\sum_{x\in \mu^{-1}Y} (x+\alpha)^{\sigma+1} = \beta^{\sigma+1}.$
Replacing $x=\mu^{-1}y$, we get
$\sum_{y\in Y} (\mu^{-1}y+\alpha)^{\sigma+1} = \beta^{\sigma+1},$
and, multiplying by $\mu^{\sigma+1}$, we obtain
$\sum_{y\in Y} (y+\mu\alpha)^{\sigma+1} = (\mu\beta)^{\sigma+1}.$
\end{proof} 

\begin{lemma}\label{l3} 
The automorphism $x\to x^2$ of the field $F$ 
maps $H_{\alpha}(\beta)$ to $H_{\alpha^2}(\beta^2)$.
\end{lemma}

\begin{lemma}\label{l4} For every $Y$ from $H$, 
the mapping $X \to X \triangle Y$ 
maps $H_\alpha(\beta)$ to 
$H_{\alpha}((\beta^{\sigma+1}+s_Y)^{1/(\sigma+1)})$ 
where $s_Y=\sum_{x\in Y}x^{\sigma+1}$. 
In particular, if $s_Y=0$, 
the mapping maps $H_\alpha^1$ to itself.
\end{lemma}
\begin{proof}
From (\ref{eq:Hab}) we have
$\sum_{x\in X\triangle Y} (x+\alpha)^{\sigma+1} 
= \sum_{x\in Y} (x+\alpha)^{\sigma+1} 
  + \beta^{\sigma+1}
= \sum_{x\in Y} x^{\sigma+1} 
  + \alpha \left(\sum_{x\in Y} x \right)^\sigma 
  + \alpha^\sigma \sum_{x\in Y} x 
  + \alpha^{\sigma+1} \sum_{x\in Y} 1 
  +  \beta^{\sigma+1}
  = s_Y+\beta^{\sigma+1},$ 
  where the last equality 
  comes from (\ref{eq:H}) 
  and from the even cardinality of $Y$.
\end{proof} 

 By an \textit{automorphism} of the partition 
 $\mathbf{H}$ 
we mean a transform $X \to \pi(X) \triangle Y$, 
where $\pi$ is a permutation of $F$ and 
$Y$ is a fixed subset of $F$, 
that maps every cell of $\mathbf{H}$ 
to another cell of $\mathbf{H}$.
The partition is called \textit{vertex-transitive} 
if the automorphism group acts transitively 
on the vertices, the odd-cardinality subsets of $F$. 
That is, for every two odd-cardinality 
subsets $X$, $Y$ of $F$ 
there is an automorphism of $\mathbf{H}$ 
that sends $X$ to $Y$.
Similarly, the \textit{cell transitivity} 
is defined (see \cite{Sol:2009:part} 
for examples of cell-transitive partitions), 
which property is, evidently, 
weaker than the vertex transitivity.
The partition is called \textit{$2$-cell-transitive} 
if the automorphism group acts transitively 
on the ordered pairs of codes from $\mathbf{H}$; 
that is, for every different 
$H'$, $H''$ from $\mathbf{H}$
and every different $H'''$, $H''''$ from $\mathbf{H}$ 
there is an automorphism of $\mathbf{H}$ 
that maps $H'$ to $H'''$ and $H''$ to $H''''$. 

\begin{proposition}\label{pro:cell}
The partition $\mathbf{H}$ of the odd-cardinality 
subsets of $F$ is $2$-cell-transitive.
\end{proposition}
\begin{proof}
By Lemmas~\ref{l1} and~\ref{l2}, 
for every different $\delta$ and $\alpha$ from $F$,
the mapping 
$M_{\delta,\alpha}:x\to (\alpha + \delta)^{-1}(x+\delta)$
maps $H^1_\delta $ and $H^1_\alpha $ 
to $H^1_0$ and $H^1_1$, respectively.
\end{proof}

\begin{proposition}\label{pro:orb}
Under the action of the automorphism group 
of $\mathbf{H}$, the set 
of odd-cardinality subsets of $F$ 
is partitioned into at most two orbits.
\end{proposition}
\begin{proof}
Since $H$ is an extended Hamming code, 
every odd-cardinality subset $X$ of $F$ 
is at distance $1$ from some $Y=Y(X)\in H$.

Let us show that $X$ and $Z$ such that 
$s_{Y(X)},s_{Y(Z)} \neq 0$ 
(in notation of Lemma~\ref{l4}) 
belong to the same orbit.
By Lemma~\ref{l4}, the mapping 
$M_X:\, X \to X \triangle Y$ 
maps every $H^1_\alpha$, $\alpha\in F$, 
to $H_{\alpha}(\beta)$ 
with $\beta = s_Y^{1/(\sigma+1)}$.
Moreover, $X$ is mapped to $\{\delta \}$ 
for some $\delta$.
By Lemma~\ref{l2}, the mapping 
$N_X:\, x\to \beta^{-1} x$ 
maps $H_{\alpha}(\beta)$ 
to $H_{\beta^{-1}\alpha}(1)$, 
while 
$N_X(M_X(X)) = N_X(\{\delta\}) = \{\beta^{-1}\delta\}$.
Finally, by Lemma~\ref{l1}, 
the mapping 
$L_X:\, x\to x+\beta^{-1}\delta$ maps 
$H_{\beta^{-1}\alpha}(1)$ to 
$H_{\beta^{-1}\alpha+\beta^{-1}\delta}(1)$, 
while $L_X(N_X(M_X(X)))=\{0\}$.
So, the mapping $L_X(N_X(M_X(\cdot)))$ 
maps $\mathbf{H}$ to $\{H_\alpha(1)\}_{\alpha\in F}$ 
and sends $X$ to $\{0\}$.
The same is true for $Z$.
In summary, 
$M_Z^{-1}(N_Z^{-1}(L_Z^{-1}(L_X(N_X(M_X(X))))))=Z$ 
and 
$M_Z^{-1}(N_Z^{-1}(L_Z^{-1}(L_X(N_X(M_X(\cdot))))))$ 
is an automorphism of $\mathbf{H}$. 
That is, $X$ and $Z$ are from the same orbit.

Similar (even more simple) arguments 
solve the case $s_{Y(X)} = s_{Y(Z)} = 0$.
\end{proof}

A computer experiment shows that for $m=5,7,9,11,13$ the partition we construct is not vertex-transitive (for $m=3$, it is); i.e., the number of the orbits is exactly $2$.
An invariant that distinguishes the vertices of different orbits is the number of two-color squares in the neighborhood:
for a given vertex $X$, we count the number $Q_X$ of quadruples $$\{X\triangle \{x,y\},X\triangle \{y,z\}, X\triangle \{z,v\}, X\triangle \{v,x\} \}$$ such that 
    $\alpha(X\triangle \{x,y\})=\alpha(X\triangle \{z,v\})$, 
and $\alpha(X\triangle \{y,z\})=\alpha(X\triangle \{v,x\})$. 
It occurs that for two vertices $X=\emptyset$ and $Y$ from different orbits, 
the numbers $Q_X$ and $Q_Y$ are different, $m\leq 13$. 
Moreover, they depend on $\sigma$, which implies that the construction, 
for fixed $m=5,7,9,11,13$, gives nonequivalent partitions.
Here is the  list of the calculated tuples $(2^m,\sigma+1,Q_X,Q_Y)$: 
$$\footnotesize
\begin{array}{|r@{}r@{\ }r@{\ }r@{\ }r|r@{}r@{\ }r@{\ }r@{\ }r|r@{}r@{\ }r@{\ }r@{\ }r|}
\hline
(& 2^5,& 3,& 155,& 115 )  & (& 2^7,& 3,& 2667,& 1995 )&  (& 2^{13},& 3,& 8412157,& 8385533 )\\
(& 2^5,& 5,& 0,& 120 )& (& 2^7,& 5,& 2667,& 1995 )&  (& 2^{13},& 5,& 8518640,& 8385520 )\\
\hhline{|-----|~~~~~|~~~~~|}
(& 2^{11},& 3,& 540408,& 523512 )& (& 2^7,& 9,& 0,& 2016 )&  (& 2^{13},& 9,& 9157538,& 8385442 )\\
\hhline{|~~~~~|-----|~~~~~|}
(& 2^{11},& 5,& 585442,& 523490 )& (& 2^9,& 3,& 36792,& 32184 )& (& 2^{13},& 17,& 7879742 ,& 8385598 )\\
(& 2^{11},& 9,& 607959,& 523479 )& (& 2^9,& 5,& 18396,& 32220 )&  (& 2^{13},& 33,& 6388980,& 8385780 )\\
(& 2^{11},& 17,& 360272,& 523600 )& (& 2^9,& 17,& 0,& 32256 )& (& 2^{13},& 65,& 0,& 8386560 )\\
\hhline{|~~~~~|-----|-----|}
(& 2^{11},& 33,& 0,& 523776 )\\ 
\hhline{|-----|~~~~~~~~~~}
\end{array}
$$
(it is sufficient to consider only the cases $\sigma< 2^{m/2}$, as the partitions for $\sigma= 2^s$ and for $\sigma= 2^{m-s}$ coincide, 
which easily follows from the identity $x^{2^s+1} = (x^{1+2^{m-s}})^{2^s}$).

{\small\textbf{Observations.} 
\textbf{1.} For $\sigma = 2^{(m-1)/2}$, we have got $Q_X=0$. 

\textbf{2.} The value $Q_Y$ is rather close to the ``average'' value $D/(2^m-1)$ 
where $D=(2^m-1)(2^{m-1}-1)2^{m-2}$ is the number of the pairs $\{X\triangle \{x,y\},X\triangle \{z,v\} \}$ such that  $\alpha(X\triangle \{x,y\})=\alpha(X\triangle \{z,v\})$.

\textbf{3.} A two-color square have never been extended to a three-color octahedron; i.e., $\alpha(X\triangle \{x,z\}) \neq \alpha(X\triangle \{y,v\})$, in the notations above.

\textbf{4.} 
The length-$2^7$ partitions with 
$ \sigma+1= 3$ and $ \sigma+1= 5$ 
are nonequivalent (while $(Q_X,Q_Y)$ coincide). 
This fact was established using 
the non-equivalence of the so-called 
local quasigroups 
$$ (F,\bullet):\ x\bullet y 
= \alpha(\{0\}\triangle\{x\}\triangle\{y\})$$
of the two partitions. 
The calculations were made using 
the isomorphism search functionality 
of the LOOPS package \cite{LOOPS} for GAP \cite{GAP}.
                                               
An evident \textbf{conjecture} is that the number of orbits is $2$ for every odd $m\geq 5$.
If this is true, then every automorphism of $\mathbf{H}$ 
is a composition of automorphisms from Lemmas 1, 2, 3, and 4 (with $s_Y=0$).
For $\sigma=2$, this follows from \cite{Berger:94} 
(any automorphism of $\mathbf{H}$ is an automorphism of the code $\{Y\in H\, : \, s_Y=0\}$,
which is an extended double-error-correcting BCH code). 

\section{Questions and remarks} \label{s:qr}
\textbf{1.}
As follows from the proof, the claim (i) of the theorem will remain valid if we replace the function $f(x)=x^{\sigma+1}$ 
by an arbitrary permutation (bijective function) $f:F\to F$
such that for all $\alpha\neq 0$ the set $H_\alpha=\{f(x)+f(x+\alpha)\,:\,x\in F\}$ 
is an affine subspace (over $\mathrm{GF}(2)$) of $F$
(the sum of an odd number of elements of $H_\alpha$ belongs to $H_\alpha$,
which does not contain  $0$ as $f$ is bijective). A class of functions that obviously satisfy this condition is the class of quadratic function,
i.e., the vector functions whose components are represented as polynomials of degree at most $2$. For such permutations, 
the cardinalities of the mutual intersections of the corresponding Hamming codes can be counted in terms of the cardinalities of $H_\alpha$,
using the technique from the second part of the theorem.
Another appropriate class of permutations is the following.
A permutations $f:F\to F$ is called {\em crooked} \cite{BenFDFla:1998} if for all $\alpha\neq 0$ the set 
$H_\alpha$ is an affine hyperplane. The Gold functions are crooked and quadratic; 
at the moment, all known crooked functions are quadratic.
Does (ii) hold for the non-quadratic crooked permutations $f(x)$ (if there are some)?
See \cite{vDamFDFla:2003} for other applications of the crooked permutations in the construction 
of different extremal combinatorial structures.

\textbf{2.} Do there exist partitions 
of the hypercube 
into cosets of maximally 
nonparallel Hamming codes
for even $m$? In \cite{HedSol:2009}, partitions 
of the space into mutually nonparallel cosets of Hamming codes are constructed for all $m\geq 3$.
The problem of minimizing mutual intersection of the Hamming codes for such 
a partition remains open for even $m$.

\section*{Acknowledgments} 
The author thanks Sergey Avgustinovich for discussions, Gohar Kyureghyan for
consulting in the area of crooked functions, and the referees for many suggestions, 
which greatly improved the presentation.
The research was partially supported
by the RFBR (grant 13-01-00463), 
by the Ministry of education and science 
of Russian Federation
(project 8227), 
and by the Target program of SB RAS for 2012-2014
(integration project No. 14).



\providecommand\href[2]{#2} \providecommand\url[1]{\href{#1}{#1}}
  \providecommand\bblmay{May} \providecommand\bbloct{October}
  \providecommand\bblsep{September} \def\DOI#1{{\small {DOI}:
  \href{http://dx.doi.org/#1}{#1}}}\def\DOIURL#1#2{{\small{DOI}:
  \href{http://dx.doi.org/#2}{#1}}}\providecommand\bbljun{June}

\end{document}